# A New Scale for Attribute Dependency in Large Database Systems

Soumya Sen[1], Anjan Dutta[2], Agostino Cortesi[3], Nabendu Chaki[4]

[1,2,4] University of Calcutta, Kolkata, India
[3] Universita Ca Foscari, Venice, Italy

{[1]iamsoumyasen@gmail.com, [2]anjanshines@gmail.com, [3]cortesi@unive.it,
[4]nabendu@ieee.org}

**Abstract.** Large, data centric applications are characterized by its different attributes. In modern day, a huge majority of the large data centric applications are based on relational model. The databases are collection of tables and every table consists of numbers of attributes. The data is accessed typically through SQL queries. The queries that are being executed could be analyzed for different types of optimizations. Analysis based on different attributes used in a set of query would guide the database administrators to enhance the speed of query execution. A better model in this context would help in predicting the nature of upcoming query set. An effective prediction model would guide in different applications of database, data warehouse, data mining etc. In this paper, a numeric scale has been proposed to enumerate the strength of associations between independent data attributes. The proposed scale is built based on some probabilistic analysis of the usage of the attributes in different queries. Thus this methodology aims to predict future usage of attributes based on the current usage.

## 1 Introduction

Success of any large database application depends on the efficiency of storing the data and retrieving the same from the database. Contemporary database applications are expected to support fast response and low turn-around time irrespective of the mediums and applications. Speeding up the query processing in distributed environment is even more challenging. Users demand high speed execution over internet, mobile phone or any other modern electrical gadgets. Fetching of closely related data attributes together would help to reduce the latency. This would be particularly significant to reduce communication cost for query processing in a distributed database. The core technology proposed in this paper could further be extended in cloud computing environment where data is distributed in different data centers. Faster query processing in cloud computing environment result in quick service processing to the users. Each of these diverse application platforms have specific, and distinct features and could be differentiated based on their nature. Thus not only the data stored in the database is subject of interest. Proper analysis of different run time parameters of the execution environment could excel the performance.

In this paper, a numeric scale has been proposed to measure the degree of association among attributes based on their usage in recent queries. This forms the foundation for several optimization aspects that could improve different database perspectives such as building materialized view, maintain indexes, formation of database clusters etc.

Attribute is the most granular form of representing data in database applications. Thus this analysis based on attributes give a deep insight of the system. Hence the deployment of optimization techniques based on this scale would help to improve the performance of the application from the granular level. Once developed, this numeric scale could be rebuilt dynamically depending on the changing nature of the queries over time. The proposed scale takes into consideration the independent characteristics of diverse applications or execution environment. Hence, by incorporating the assumptions and constraints of the specific system, this scale could be used in heterogeneous applications.

The rest of this paper is organized in several sections after this brief introduction. Section 2 describes different existing work on optimization of query execution. Section 3 contains the proposed methodology of constructing the numeric scale. In section 4, the selection of parameters is discussed along with the complexity analysis of this method. In section 5, the entire process is illustrated through an example. The concluding remarks in section 6 summarize the work and mention the future extensions and applications of the proposed methodology.

## 2 Related Work

Optimizing the query processing in large data centric application has traditionally been studied under the name of query optimization. Several works has been reported on this area. Initially the focus was on simple database. Over the time, multiple aspects including diverse performance criteria and constraints as well as the requirements for specific applications are taken into consideration. In the rest of this section, a brief survey work has been presented focused on query optimization.

A method of query processing and standardization is proposed in [1] where query graphs are used to represent the queries. These are converted to query trees which in turn are represented in canonical vector form. These trees are optimized to enhance the performance. Another graph based model used in this context help in further optimization by considering the parameters like relation size, tuple size, join selectivity factors [2]. An algorithm has also been proposed in [2] to find a near optimal execution plan in polynomial time. The time complexity of analyzing the graph is often quite high. In [2], instead of considering the whole execution path, selective paths are processed. Genetic algorithm and heuristic approaches are also used for query optimization [3, 4]. A genetic algorithm [3] to minimize the data transmission cost required for a distributed set query processing is presented. This work is also a contribution in the distributed database application. On the other hand, a heuristic approach [4] is proposed to efficiently derive execution plans for complex queries. These works takes into account presence of index and goes beyond simple join reordering. Mathematical model is also helpful in this context. Tarski Algebra [5] along with graphical representation of query is used to achieve efficient query optimization. Another graph based approach is shown to optimize the linear recursive queries [6] in SQL. This approach computes transitive closure of a graph and computes the power matrix of its adjacency matrix. Using this [6], optimization plan is evaluated for four types of graphs: binary tree, list, cyclic graph and complete graph. In recent past, the distributed query optimization get serious research attention as many of the current applications run in distributed environment. A multi-query optimization aspect for distributed similarity query processing [7] attempts to exploit the dependencies in the derivation of a query evaluation plan. A four-step algorithm [7] is proposed to minimize the response time and towards increasing the parallelism of I/O and CPU.

Creation of materialized view and its archival in fast cache also helps in reducing the query access time by prior assessment of the data that is frequently accessed. The view management process would be more effective if such data can be included in the view that is likely to be accessed in near future. An algorithm was been proposed by Yang, et. al. that utilizes a Multiple View Processing Plan (MVPP) [8] to obtain an optimal materialized view selection. The objective had been to achieve the combination of good performance and low maintenance cost. However, Yang's approach did not consider the system storage constraints. Gupta proposed a greedy algorithm [9] to incorporate the maintenance cost and storage constraint in the selection of data warehouse materialized views. The AND-OR view graphs were used [9] to represent all possible ways to generate warehouse views such that the best query path can be utilized to optimize query response time.

Materialized views are built to minimize the total query response time while there is an associated overhead towards the creation and maintenance of these views. Thus, an effort has always been to balance a strike between optimizing the processing cost [10] and time for view selection vis-à-vis increasing the efficiency of query processing [11] by utilizing well-organized materialized view.

Use of index also helps to achieve faster data processing. The application of bitmap index [12] helps in query processing for data warehouse and decision-making systems. The work proposed in [13], couples the materialized view and index selection to take view-index interactions into account and achieve efficient storage space sharing.

However, none of these existing works of query optimization and query processing is based on the outcome of a quantitative analysis on the intensity of association between the attributes. The work proposed in this paper, therefore, aims towards finding a numeric measure to assess such inter-attribute associations. This numeric scale provides the relationship between attributes in terms of both present and future usage. Hence this scale could be applicable in any optimization methods that involve the association of a set of attributes. The knowledge of this scale could be used in building and maintenance of materialized views or indexes.

## 3  Attribute Scaling

The motivation behind this work is to create a numeric scale to represent the degree of associations between different attributes based on a set of queries. This scale would help to generate different materialized views based on the requirement of the users. The proposed methodology of constructing this numeric scale is explained using a seven step algorithm named Numeric_Scale (described in section 3.1).  In the pre-processing phase, a set of queries (say m number of queries) are picked from the recent queries evoked in an application. Say, a total of n numbers of attributes participate in the query set. The proposed scale is based on these n attributes.

The 1st step of the proposed algorithm builds the Query Attribute Usage Matrix (QAUM), which shows what attributes are used by which queries (described in section 3.2). In the next step, mutual dependencies among every pair of attributes are computed, yielding to the Attribute Dependency Matrix (ADM) (described in 3.3). This is a symmetric matrix. Based on the result of ADM a new matrix, called Probability Distribution Matrix (PDM), is computed. PDM shows the dependencies among every pair of attributes based on a probabilistic function (described in 3.4). This is followed by the computation of standard deviation of each attribute (described in section 3.5).  Then for every attribute, scaling is calculated using a function of standard deviation and frequency of attribute occurrences (described in section 3.6). This result is stored in the Numeric Scale Matrix (NSM).  Now this result is normalized in a scale of 10 for every attribute and stored in Normalized Numeric Scale Matrix (NNSM) (described in section 3.7). The NNSM Matrix shows the dependency among all pair of attributes in the query set based on a numeric scale. Higher the value in each cell of NNSM lower the dependency among the pair of attributes corresponding to the particular cell. Thus the entry of 10 in some cell say, [i, j] means that for ith attribute it has lowest dependency on jth attribute.

### 3.1  Algorithnm Numeric_Scale

**Begin**
**Step 1.** The association between the queries and attributes is computed in Query Attribute Usage Matrix (QAUM).
   Call method QAUM_Computation;
**Step 2.** Mutual dependencies of the attributes are stored in Attribute Dependency Matrix (ADM). The sum of 1 to nth columns (except the diagonal cell) for a given tuple is stored in the newly inserted $(m+1)^{th}$ column of ADM known as Total Measure.
   Call method ADM_Computation;
**Step 3.** The probability that an attribute is dependent on another attribute is calculated and stored in a Probability Distribution Matrix (PDM).
   Call method Probability_Calculation;
**Step 4.** Standard Deviation (SD) of each attribute is calculated.
   Call  method StandardDeviation_Computation;

**Step 5.** A particular attribute (PIVOT attribute) is selected and scaling of each attribute is done using the methodology Scaling_Calculation and the result is stored in Numeric Scale Matrix [NSM].
   Call method Scaling_Calculation;
**Step 6.** Normalize the computed value of NSM in the closed interval of [1, 10] and stored in Normalized Numeric Scale Matrix[ NNSM].
   Call method Normalized_Scale;
**End Numeric_Scale.**

### 3.2 Method QAUM_Computation

In this stage, a m x n binary valued matrix is constructed named as Query Attribute Usage Matrix (QAUM). Here, m is the numbers of queries in the query set and n is the total numbers of attributes used in this query set. If query h uses $k^{th}$ attribute, QAUM[h, k] would be 1 else 0.

**Begin QAUM_Computation**
/* Procedure to build Query Attribute Usage Matrix (QAUM) */
   $\forall h \; \epsilon \; [1..m], \forall k \; \epsilon \; [1..n], if \; k \; is \; used \; in \; h,$
       $QAUM_{h,k} = 1;$
   else
       $QAUM_{h,k} = 0;$
**End QAUM_Computation**

### 3.3 Method ADM_Computation

In this stage, a n x n symmetric matrix named Attribute Dependency Matrix (ADM) is built. Each cell say [h, k] of this matrix keeps a count on the number of times that both $h^{th}$ and $k^{th}$ attributes are used simultaneously in the set of m queries. As this is a symmetric matrix at this stage ADM[h, k] = ADM[k, h]. The diagonal of this matrix is marked as '#'. The diagonal cells contain trivial information that the dependency of an attribute is with itself only. After this new column is inserted into ADM named Total Measure, which stores the sum of every row. So, finally ADM is a n x (n+1) matrix.

**Begin ADM_Computation**
/* Procedure to count number of times two attributes a, b occur simultaneously and store it in matrix ADM and finally adding the values of each row to store in column Total Measure.*/
   $\forall h, k \; \epsilon \; [1..n], if \; h = k$
       $ADM_{h,k} = \#;$
   else
       $ADM_{h,k} = total \; count \; of \; occurences \; of$
       $attributes \; h \; and \; k \; together \; in \; the \; set \; of \; N \; queries \;;$
   $\forall h, k \; \epsilon \; [1..n],$
       $ADM_{h,k+1} = {}_{k=1}^{n}\sum ADM_{h,k} \; \forall \; h \neq k.$
**End ADM_Computation**

### 3.4 Method Probability_Calculation

In this stage an n x n Probability Distribution Matrix (PDM) is constructed. This matrix is build to estimate a probabilistic measure of dependencies of every $h^{th}$ attribute with other attributes. Every value of PDM[h, k] is computed by dividing the value of ADM[h, k] by the value of Total Measure (ADM[h, n+1]) corresponding to the $h^{th}$ row of ADM. However, computing the measures of two types of cells are not required. These are diagonal cells and the cells for which ADM entry is 0. These types of cells are marked as '#' in PDM

**Begin Probability_Calculation**
/* Procedure to build Probability Distribution Matrix (PDM) on the basis of use of attributes */
$\forall h, k \in [1..n], if\ (h = k) \vee (ADM_{h,k} = 0),$
$\quad PDM_{h,k} = \#;$
$else$
$\quad PDM_{h,k} = \frac{ADM_{h,k}}{ADM_{h,n+1}};$
**End Probability_Calculation**

### 3.5 Method StandardDeviation_Computation

In this stage the mean, variance and standard deviation of attributes are computed as function of ADM and PDM. This is computed to measure the deviation of mean of other attribute from a given attribute.

$$\text{Mean}(\mu) = {}_{h=1}^{n}\Sigma p_h.x_h$$

$$\text{Variance}(X) = {}_{h=1}^{n}\Sigma p_h.(x_h - \mu)^2$$

$$\text{Standard Deviation(SD)} = \sqrt{\text{Variance}(X)}$$

**Fig.1.** Formulas for Mean, Variance and Standard Deviation.

If the random variable X is discrete with probability mass function $x_1 \rightarrow p_1, \ldots, x_n \rightarrow p_n$ then Mean, Variance and Standard Deviation(SD) are calculated using the three formulas shown in Figure. 1. Here, $p_h$ and $x_h$ are the entries of PDM and ADM respectively. However those entries which are marked as # in PDM they are not considered in this computation. The results of Mean, Variance and Standard Deviation of every attribute are stored in MVSD table. The $1^{st}$ row contains the mean, $2^{nd}$ row contains the variance and $3^{rd}$ row contains the standard deviation.

**Begin StandardDeviation_Computation**
/* Procedure to compute and store mean, variance and standard deviation for attributes in MVSD matrix*/
$\forall h \in [1..n]$

```
    S = 0;
    ∀k ∈ [1..n] if (PDM_{h,k} ≠ #)
         S = S + ADM_{h,k} × PDM_{h,k};  /*Value stored in ADM [h, k] is multiplied with
         the value stored in PDM[h,k] */
MVSD_{1,h} = S;    /* Stores mean(μ) */
∀h ∈ [1..n]
    SD = 0;
    ∀k ∈ [1..n] if (PDM_{h,k} ≠ #)
         SD = SD + PDM_{h,k} × (ADM_{h,k} − MVSD_{1,h})²;
MVSD_{2,h} = SD;  /* Stores Variance(X) */
MVSD_{3,h} = √SD;            /* Stores Standards Deviation */
```
**End StandardDeviation_Computation**

### 3.6 Method Scaling _Calculation

In this stage an n x n matrix is constructed and named as Numeric Scale Matrix (NSM). The values of this matrix are computed as the function of standard deviation in MVSD and ADM. The result of every MVSD[h, k] is computed as : modulus difference of standard deviation of $h^{th}$ and $k^{th}$ attribute, which is divided by the ADM[h,k]. However, if the PDM[h, k] is #, it is not considered for computation. This matrix is constructed taking the help of both the probabilistic estimate of attribute usage as well as the current context of attribute usage. Thus this matrix identifies the degree of interdependence among every pair of attribute. Lower the value in every cell of NSM, higher the degree of dependence among the attributes corresponds to the row and column.

**Begin Scaling_calculation**
   /* Procedure to compute and store degree of interdependence among attributes to build Numeric Scale Matrix (NSM)*/

```
    ∀h ∈ [1..m], ∀k ∈ [1..n], if PDM_{h,k} = #,
         NSM_{h,k} = #;
    else
         D = ADM_{h,k};
         NSM_{h,k} = |MVSD_{3,h} − MVSD_{3,k}| / D;
```
**End Scaling_Calculation**

### 3.7 Method Normalized_Scale

In this stage another n x n matrix named Normalized Numeric Scale Matrix (NNSM) is constructed by normalizing every row of NSM in a scale of 10. For every row the highest value is mapped to 10, similarly all other values of the row are mapped to the new value with the same mapping function. For a row (say, for attribute h) if the attribute k has the value 10, that means h has the weakest relationship with attribute k where as the lowest entry (say, for attribute p) in some column signifies that h has the strongest relationship with attribute p.

**Begin Normalized_Scale**
/* Procedure to compute normalized numeric scale and result stored in NNSM*/
$\forall h \in [1..n]$
   $Max = 0;$
   $\forall k \in [1..n] \; if \; (NSM_{h,k} \neq \#) \wedge (NSM_{h,k} > Max)$
      $Max = NSM_{h,k};$
$\forall k \in [1..n] \; if \; (NSM_{h,k} \neq \#)$
      $NNSM_{h,k} = \#;$
   else
      $NNSM_{h,k} = (NSM_{h,k}/Max) \times 10;$
**End Normalized_Scale**

## 4  Selection of Parameters and Complexity Analysis

The proposed model is based on set of queries and the attributes belonging to this set. Thus some selection criterions are important for successful execution of it. The different performance issues for the proposed numeric scale include scalability, dynamicity, and generalization aspect. The roles of the identified parameters are discussed below:

1) Query Selection: This algorithm starts with a set of queries. The entire analysis process is based on this query set. Hence, identification of query set is an important parameter for this process. It could be done in different ways. Two of the widely used methods are random selection, and interval based Selection. The first method extracts some of the executed queries from a given set randomly. In the second approach, certain time interval is chosen and the queries that have been executed during this are taken for analysis purpose.

2) Attribute Selection: Once the queries have been selected a set of attributes belonging to this query set is clearly identified. However, all of the attributes may not be subject of interest. As for example, if an attribute is used rarely in a query set, discarding that attribute would reduce the size of ADM and hence result in a faster execution of this method.

3) Threshold Selection: In the preceding step the requirement of attribute selection is defined. This is to be supported by some proper usage ratio. Thus the selection of threshold value of usage is also need to be defined.

   The overall asymptotic run-time complexity of this algorithm is $O(n^2)$, where n is the number of attributes selected for analysis. Therefore, the effectiveness of our approach relies on one hand on the ratio n/M, where M is the overall number of attributes in the database, and on the other hand variance degree of attributes appearing in the query sequence. Measuring the actual computational advantage of our algorithm is the main subject of our ongoing work.

## 5  Illustrative Example

Let's consider a small example set of queries. This is only for the sake of a lucid explanation of the steps to be followed in the proposed algorithm. There are ten queries (q1, q2, …. ,q10) in the set which use ten different attributes namely a1, a2,……,a10. The queries are not given here due to space constraint, the example is shown starting from QAUM. The results are shown up to 2 decimal places.

Step 1: The use of these 10 attributes, by these 10 queries is shown in the QAUM(Table 1) using the method QAUM_Computation. If we consider query q1 we can say this query uses attributes a1, a2, a3, a4, a5 and a9.

**Table 1.**  QAUM.

|     | a1 | a2 | a3 | a4 | a5 | a6 | a7 | a8 | a9 | a10 |
|-----|----|----|----|----|----|----|----|----|----|-----|
| q1  | 1  | 1  | 1  | 1  | 1  | 0  | 0  | 0  | 1  | 0   |
| q2  | 1  | 0  | 0  | 1  | 0  | 1  | 1  | 0  | 0  | 0   |
| q3  | 0  | 0  | 1  | 0  | 1  | 1  | 1  | 1  | 0  | 0   |
| q4  | 1  | 0  | 0  | 0  | 1  | 1  | 0  | 0  | 1  | 1   |
| q5  | 0  | 1  | 0  | 0  | 0  | 0  | 1  | 1  | 1  | 1   |
| q6  | 0  | 0  | 1  | 1  | 0  | 1  | 0  | 0  | 0  | 1   |
| q7  | 1  | 1  | 1  | 0  | 0  | 1  | 0  | 1  | 1  | 0   |
| q8  | 1  | 1  | 1  | 0  | 1  | 0  | 0  | 0  | 0  | 1   |
| q9  | 0  | 1  | 1  | 0  | 1  | 0  | 0  | 1  | 1  | 1   |
| q10 | 1  | 0  | 0  | 1  | 0  | 1  | 1  | 0  | 1  | 1   |

Step 2: The mutual dependencies among all the attributes are stored in Attribute Dependency Matrix (ADM). For example the attributes a1 and a2 are used simultaneously in three queries, namely q1, q7 and q8. Thus, the entry in ADM for (a1, a2) is 3. The Total Measure is computed in ADM by adding the attribute dependency in every row. For instance, the Total Measure for a1 is 24. (Table 2)

**Table 2.**  ADM.

|     | a1 | a2 | a3 | a4 | a5 | a6 | a7 | a8 | a9 | a10 | Total Measure |
|-----|----|----|----|----|----|----|----|----|----|-----|---------------|
| a1  | #  | 3  | 3  | 3  | 2  | 3  | 2  | 1  | 4  | 3   | 24            |
| a2  | 3  | #  | 4  | 1  | 3  | 2  | 1  | 3  | 4  | 3   | 24            |
| a3  | 3  | 4  | #  | 2  | 4  | 3  | 2  | 3  | 3  | 2   | 26            |
| a4  | 3  | 1  | 2  | #  | 1  | 3  | 2  | 0  | 2  | 2   | 16            |
| a5  | 3  | 2  | 4  | 1  | #  | 2  | 1  | 2  | 3  | 3   | 21            |
| a6  | 3  | 2  | 3  | 3  | 2  | #  | 3  | 2  | 3  | 3   | 24            |
| a7  | 2  | 1  | 2  | 2  | 1  | 3  | #  | 2  | 2  | 2   | 17            |
| a8  | 1  | 3  | 3  | 0  | 2  | 2  | 2  | #  | 3  | 2   | 18            |
| a9  | 4  | 4  | 3  | 2  | 3  | 3  | 2  | 3  | #  | 4   | 28            |
| a10 | 3  | 3  | 2  | 2  | 3  | 3  | 2  | 2  | 4  | #   | 24            |

Step 3: PDM is built (Table 3) using the method Probability_Calculation to define the probabilistic estimate of attribute occurrence; e.g., (a1, a2) in PDM is computed by dividing ADM(a1,a2) with the Total Measure of a1 from ADM.

**Table 3.** PDM.

|     | a1   | a2   | a3   | a4   | a5   | a6   | a7   | a8   | a9   | a10  |
|-----|------|------|------|------|------|------|------|------|------|------|
| a1  | #    | 0.13 | 0.13 | 0.13 | 0.08 | 0.13 | 0.08 | 0.04 | 0.17 | 0.13 |
| a2  | 0.13 | #    | 0.17 | 0.04 | 0.13 | 0.08 | 0.04 | 0.13 | 0.17 | 0.13 |
| a3  | 0.12 | 0.15 | #    | 0.08 | 0.15 | 0.12 | 0.08 | 0.12 | 0.12 | 0.08 |
| a4  | 0.19 | 0.06 | 0.13 | #    | 0.06 | 0.19 | 0.13 | #    | 0.13 | 0.13 |
| a5  | 0.14 | 0.10 | 0.19 | 0.05 | #    | 0.10 | 0.05 | 0.10 | 0.14 | 0.14 |
| a6  | 0.13 | 0.08 | 0.13 | 0.13 | 0.08 | #    | 0.13 | 0.08 | 0.13 | 0.13 |
| a7  | 0.12 | 0.06 | 0.12 | 0.12 | 0.06 | 0.18 | #    | 0.12 | 0.12 | 0.12 |
| a8  | 0.06 | 0.17 | 0.17 | 0.00 | 0.11 | 0.11 | 0.11 | #    | 0.17 | 0.11 |
| a9  | 0.14 | 0.14 | 0.11 | 0.07 | 0.11 | 0.11 | 0.07 | 0.11 | #    | 0.14 |
| a10 | 0.13 | 0.13 | 0.08 | 0.08 | 0.13 | 0.13 | 0.08 | 0.08 | 0.17 | #    |

Step 4: Using the method StandardDeviation_Computation Mean(μ), Variance, standard deviation for all the attributes are computed and stored in table MVSD (Table 4). It has three data rows and n columns for the attributes. The first row of the table contains the mean, the second row contains variance and the third row contains the standard deviation (SD). The formulations for these three counts are specified in Fig. 1.

**Table 4.** MVSD

|          | a1   | a2   | a3   | a4   | a5   | a6   | a7   | a8   | a9   | a10  |
|----------|------|------|------|------|------|------|------|------|------|------|
| Mean     | 2.92 | 3.08 | 3.08 | 2.25 | 2.71 | 2.75 | 2.06 | 2.44 | 3.29 | 2.83 |
| Variance | 0.30 | 2.09 | 2.61 | 5.54 | 3.95 | 3.82 | 3.46 | 3.41 | 6.98 | 7.38 |
| SD       | 0.55 | 1.45 | 1.62 | 2.35 | 1.99 | 1.95 | 1.86 | 1.85 | 2.64 | 2.72 |

**Table 5.** NSM

|     | a1   | a2   | a3   | a4   | a5   | a6   | a7   | a8   | a9   | a10  |
|-----|------|------|------|------|------|------|------|------|------|------|
| a1  | #    | 0.30 | 0.36 | 0.60 | 0.72 | 0.47 | 0.66 | 1.30 | 0.52 | 0.72 |
| a2  | 0.30 | #    | 0.04 | 0.90 | 0.18 | 0.25 | 0.41 | 0.13 | 0.30 | 0.42 |
| a3  | 0.36 | 0.04 | #    | 0.37 | 0.09 | 0.11 | 0.12 | 0.08 | 0.34 | 0.55 |
| a4  | 0.60 | 0.90 | 0.37 | #    | 0.36 | 0.13 | 0.25 | #    | 0.15 | 0.19 |
| a5  | 0.48 | 0.27 | 0.09 | 0.36 | #    | 0.02 | 0.13 | 0.07 | 0.22 | 0.24 |
| a6  | 0.47 | 0.25 | 0.11 | 0.13 | 0.02 | #    | 0.03 | 0.05 | 0.23 | 0.26 |
| a7  | 0.66 | 0.41 | 0.12 | 0.25 | 0.13 | 0.03 | #    | 0.01 | 0.39 | 0.43 |
| a8  | 1.30 | 0.13 | 0.08 | 0.00 | 0.07 | 0.05 | 0.01 | #    | 0.26 | 0.44 |
| a9  | 0.52 | 0.30 | 0.34 | 0.15 | 0.22 | 0.23 | 0.39 | 0.26 | #    | 0.02 |
| a10 | 0.72 | 0.42 | 0.55 | 0.19 | 0.24 | 0.26 | 0.43 | 0.44 | 0.02 | #    |

Step 5: Using the method Scaling_Calculation, the NSM table (Table 5) is constructed. The entries in NSM are derived from the corresponding entries in ADM and MVSD. As for example NSM(a1, a2) is computed at first by taking the modulo subtraction result of standard deviation of a1 and a2. Then this result is divided by ADM (a1, a2). In this case, the difference from the modulo subtraction is 0.90 and ADM(a1, a2) is 3. Thus the NSM(a1, a2) is 0.30.

Table 6. NNSM.

|     | a1   | a2   | a3   | a4   | a5   | a6   | a7   | a8   | a9   | a10  |
|-----|------|------|------|------|------|------|------|------|------|------|
| a1  | #    | 2.31 | 2.74 | 4.62 | 5.54 | 3.59 | 5.04 | 10   | 4.02 | 5.56 |
| a2  | 3.33 | #    | 0.47 | 10   | 2.00 | 2.78 | 4.56 | 1.48 | 3.31 | 4.70 |
| a3  | 6.48 | 0.77 | #    | 6.64 | 1.68 | 2.00 | 2.18 | 1.39 | 6.18 | 10   |
| a4  | 6.67 | 10   | 4.06 | #    | 4.00 | 1.48 | 2.72 | #    | 1.61 | 2.06 |
| a5  | 10   | 5.63 | 1.93 | 7.50 | #    | 0.42 | 2.71 | 1.46 | 4.51 | 5.07 |
| a6  | 10   | 5.36 | 2.36 | 2.86 | 0.43 | #    | 0.64 | 1.07 | 4.93 | 5.50 |
| a7  | 10   | 6.26 | 1.83 | 3.74 | 1.98 | 0.46 | #    | 0.08 | 5.95 | 6.56 |
| a8  | 10   | 1.03 | 0.59 | #    | 0.54 | 0.38 | 0.04 | #    | 2.03 | 3.35 |
| a9  | 10   | 5.69 | 6.51 | 2.78 | 4.15 | 4.40 | 7.46 | 5.04 | #    | 0.38 |
| a10 | 10   | 5.85 | 7.60 | 2.56 | 3.36 | 3.55 | 5.94 | 6.01 | 0.28 | #    |

Step 6: Using the method Normalized_Scale, every row of NSM is scaled in a factor of 10. For every row the highest value of NSM is scaled to 10 and similarly all other attributes are mapped to new values in NNSM(Table 6). As for example in the 1$^{st}$ row of NSM a8 has the maximum value thus it is scaled to 10 using the algorithm described in section 3.7. Similarly all other values of 1$^{st}$ row are mapped.

## 6  Conclusion

The novelty of this paper is in proposing a methodology to build a numeric scale based on quantitative analysis on the set of attributes used in recent queries. Use of the standard deviation in this methodology helps to build a predictive model on future usage of attributes. Thus this method combines the actual usage with the probabilistic assumptions.

   The proposed scale would find significant usage in diverse aspects of database management. This would improve the performance towards creation and maintenance of the materialized views. This in turn would enhance the query execution in both database and data warehouse applications. As the proposed scale is independent of any external parameters, materialized views could be formed for heterogeneous applications. Other database functionalities like indexing, cluster formation, etc. could also be done on the basis of quantitative measures using the proposed scale as compared to intuitive approaches. The proposed scale is also useful in any rank based analysis of attributes. The future research work of this scale includes several aspects. Firstly, the types of queries to be selected to initiate this process for different applications are interesting and depend on the business logic. Experimental findings on diverse database applications by using the proposed scale could unearth interesting associations. Secondly, incorporating value based analysis over the attributes based analysis could be one using the scale. As all the values of the attributes are not accessed during query processing filtering could be used on the values as well. Combining the value based analysis with the existing numeric scale would help to achieve high speed query processing. Besides, the proposed scale could be combined with the concept of abstraction of attributes using concept hierarchy. This would help to reduce the amount of data to be accessed and to reduce size of materialized views.